\documentclass[aps,prl,reprint,twocolumn]{revtex4}

\usepackage{epsfig,amssymb,amsmath}

\usepackage{color}

\def\labell#1{\label{#1}}
\def\>{\rangle}\def\<{\langle}

\def\togli#1{}

\begin{document}


\title{A fundamental problem in quantizing general relativity} 
\author{  Lorenzo Maccone}
\affiliation{\vbox{Dip.~Fisica and INFN Sez.~Pavia, University~of
    Pavia, via Bassi 6, I-27100 Pavia, Italy} \\ }
\begin{abstract}
  We point out a fundamental problem that hinders the quantization of
  general relativity: quantum mechanics is formulated in terms of {\em
    systems}, typically limited in space but infinitely extended in
  time, while general relativity is formulated in terms of {\em
    events}, limited both in space and in time. Many of the problems
  faced while connecting the two theories stem from the difficulty in
  shoe-horning one formulation into the other. A solution is not
  presented, but a list of desiderata for a quantum theory based on
  events is laid out.
\end{abstract}


\maketitle  
Consider the axiomatic
formalization of the two main theories in modern physics, quantum
mechanics and general relativity.  Strikingly, one finds that they
refer to different objects, a fact that, perhaps, has not been widely
appreciated. Indeed, quantum mechanics deals with systems whereas
general relativity deals with events. These are substantially
different objects: a system has finite spatial extent typically, but
infinite temporal extent, whereas an event has finite spatial and
temporal extent. Clearly one may define an event as ``something that
happens to a system'', but that still considers a system as the
fundamental object, whereas a theory of spacetime should be only
formulated in terms of events and the system should be a derivative
notion: a ``succession of events''. In other words, there is a
difference between a general-relativistic theory of quantum mechanics
and a quantum theory of general relativity. The first is formulated in
terms of systems, the second in terms of events. It seems that only
the first avenue has been explored so far, without yet achieving a
definitive theory.

In this paper we detail this problem and present a list of
characteristics that a quantum theory for events could have. This is
not a solution to the problem but, hopefully, a small first step.

The formulation of quantum mechanics can be summarized by its four
axioms: (a)~The state of a system is described by a unit vector
$|\psi\>$ in a complex Hilbert space, and the system's observable
properties are described by self-adjoint operators on that space;
(b)~The state space of a composite system is given by the tensor
product of the spaces of the component systems; (c)~The time evolution
of an isolated system is described by a unitary operator acting on the
system state, $|\psi({t})\>=U_{t}|\psi({t}=0)\>$ or, equivalently, by
the Schr\"odinger equation; (d)~The probability that a measurement of
a property $A$, described by the operator with spectral decomposition
$\sum_aa\Pi_a$, returns a value $a$ is given by
$p(a)=\<\psi|\Pi_a|\psi\>$ (Born rule), where $|\psi\>$ is the
system's state. The rest of quantum theory can be derived from these
axioms.

The centrality of the concept of ``system'' is clear, it appears in
all axioms.  It is also clear that such system is eternal, namely
infinitely extended in time, since there is no prescription for any
time evolution of the Hilbert spaces \cite{unruhbook}: only the state
evolves in time (in the Schr\"odinger picture used here). While
infinitely extended in time, the system is typically not infinitely
extended in space.  There is a further asymmetry in the spacetime
description of quantum systems: time appears in the evolution
postulate (c) as a (classical) parameter $t$ which is external to the
theory, not a dynamical variable. In contrast, the spatial degrees of
freedom can typically be described as quantum properties. In other
words, time is a post-selected (classically conditioned) quantity
\cite{qtime}: $|\psi({t})\>$ describes the state {\em conditioned} on
the time being $t$ (Working in the Heisenberg picture\footnote{The
  equivalence between the Heisenberg and Schr\"odinger pictures
  assumes the existence of a Hamiltonian with some regularity
  conditions \cite{diracsch} and assumes that the Hamiltonian does not
  have nontrivial dependence on the system's parameters
  \cite{franson}.}  does not change the substance: the conditioning is
shifted to the observables.) Summarizing, a quantum system is
spacetime-asymmetric in two ways: both in its spacetime extension and
in the fact that its spatial properties are described as
quantum-numbers, while its temporal properties are described as
c-numbers.

Incidentally, both problems are bypassed in quantum field
theory.  It considers systems that are infinitely extended also in
space (fields), thus recovering the symmetry for the spacetime
extension: fields are infinitely extended both in space and in time.
It also considers space and time both as c-numbers and not dynamical
variables.  Nonetheless, these tricks are not sufficient to fully
quantize general relativity \cite{birrell} (arguably because
properties of infinitely-extended systems are ill defined in general
relativity and because field operators are evolved by a global time
coordinate\footnote{``There is no way of defining a relativistic
  proper time for a quantum system which is spread all over space''
  \cite{peresmf}.} which is meaningful only for few types of
spacetimes).

We will not remind here the axiomatic formulation of classical general
relativity (e.g.~\cite{mtw}, ch.17): for our aims it is
sufficient to remark that the theory relates the spacetime geometry to
its energy-momentum content. The object of the theory is then
spacetime, namely, events.

The tension between the two formulations is evident, and its origin is
clear: quantum mechanics is formulated in terms of systems, general
relativity in terms of events. One may argue that, in light of the
spectacular experimental successes of quantum mechanics, which has
been tested to better precision than general relativity, one should
try to retain the former and modify the latter. This has (loosely)
been the road followed up to now by the most promising ways to
quantize general relativity. Here we argue, instead, that a
modification of quantum mechanics to accommodate events might also be
a viable route. This does not necessarily entail a modification of the
formulation (a-d) introduced above. For example, it has been shown
that quantum theory can be extended to treat time as a dynamical
variable \cite{qtime,paw,ak,morse,altri,rov} in such a way that the
conventional formulation is recovered by conditioning the states and
observables on a measurement of time. A quantum mechanics formulated
in terms of events, instead of systems, might be obtained analogously.

A truly quantum description of events should be able to
assign a {\em probability amplitude for an event to happen in some
  location at some time}.  In such a theory, one would have to provide
predictions for events such as ``at time $t$, in position $x$ {\em a}
spin was found in the state $|{\uparrow}\>$''. The current quantum
physics (excluding quantum field theory) is only able to provide
predictions for properties of a system (e.g.~a particle with spin),
such as the probability that ``at time $t$, in position $x$, {\em the}
spin was found in the state $|{\uparrow}\>$'', a small but crucial
difference.  Indeed, the starting point of the theory is the
definition of the Hilbert space for the system (the particle with
spin), and then one can calculate its properties at given positions
and times. The current theory is not able to introduce a Hilbert space
for events, but only for systems.  This denotes a schizophrenic
attitude: the spatial position of a system is a quantum property
(``finding the particle at position $x$ given that time is $t$''), but
the temporal position of an event is not (``finding the particle at
time $t$ given that the position is $x$''). Indeed, if a system
(e.g.~a particle) is prepared at position $A$ at time ${t}_A$ and
detected later at position $B$ at time ${t}_B$ the statement that the
``particle crosses intermediate positions'' leads to all kinds of
double-slit-type paradoxes \cite{ahaparad} and is carefully avoided in
the conventional theory.  However, the conventional theory tells us
that the ``particles crosses all intermediate times'' between ${t}_A$
and ${t}_B$. Technically, this is implicit in the wavefunction
normalization at all times $\int dx|\<x|\psi({t})\>|^2=1\:\forall t$,
with $|x\>$ position eigenstate: we cannot assign an intermediate
position, but we know that at all intermediate times the particle
``exists'', i.e.~if one were to look, it would be found somewhere with
certainty.  This observation is just a
restatement of the fact (noted above) that positional degrees of
freedom $|x\>$ are treated dynamically, but temporal degrees of
freedom $t$ are treated parametrically. But this is not just a problem
of how the evolution postulate (c) is formulated, it is a fundamental
problem of how the theory is formulated in terms of eternal systems
that ``exist'' at all times (Fig.~\ref{f:fig}).  

{\begin{figure}[hbt]
\begin{center}
\epsfxsize=.85\hsize\leavevmode\epsffile{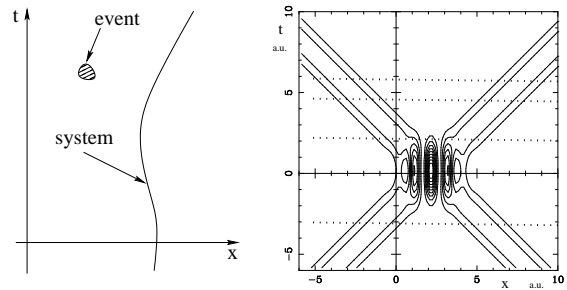}
\end{center}
\vspace{-.5cm}
\caption{Left: classical spacetime representation of an event and of a
  system (through its worldline). A classical system can be seen as a
  succession of events. Right: `worldline' of a quantum system
  (contour plot of the square modulus of the wavefunction of a
  particle spread into two counterpropagating Gaussian wavepackets).
  Its normalization at all times (horizontal dotted lines) implies
  that the system ``exists'' at all times. Namely, interference
  (superposition) in time is impossible, in contrast to interference
  in space (pictured). From general principles one expects that it
  should be possible to place quantum events in superpositions of
  different times. The current quantum formalism, based on systems
  (pictured), cannot describe this.}
\labell{f:fig}\end{figure}
}

An immediate difficulty is encountered in devising a quantum theory
for events: the identification of what is a `quantum event'. We should
not consider it as a primitive intuitive notion: it is far from
intuitive. We also cannot appeal to the textbook definitions of
`event' since they are useless. Basically two definitions appear in
the literature. The first defines `event' as a point in spacetime,
where spacetime is the set of events, e.g.~\cite{wald,schutz}. The
obvious circularity of such definition implies that it is taken as a
primitive, undefined notion.  The second defines `event' as an
intersection of world-lines, e.g.~\cite{mtw}, which provides a
physical content to the definition in order to bypass the `hole
argument'.  Unfortunately, we cannot use it in the quantum realm,
where world-lines (trajectories) are meaningless.  Nonetheless, a
closely related concept, spacetime coincidences, has been proposed
\cite{rov} in the quantum gravity context. The notion of `quantum
event' connected to it is a `joint eigenstate of a complete set of
kinematical observables' (\cite{rov}, ch.5.2.1). While a good
definition for the aims of quantum gravity, it is still based on
properties of quantum {\em systems} and it does not reduce to a
classical event in the classical limit of the Ehrenfest
theorem\footnote{Indeed its interpretation becomes unclear if one
  uses, as a complete set of observables, the momenta of all the
  particles. Even the limitation to position observables will not help
  in the case of multiple systems: it would identify multiple points in
  spacetime, the positions of particles conditioned on what a clock
  shows.  }.

We now list the properties that a quantum theory for events could
possess. We emphasize again that we do not advocate abandoning the
conventional quantum formulation (a-d) given above: that formulation
{\em must} be retained in the appropriate limits, e.g.~the case in
which experiments are timed with an external (classical) clock.
\begin{enumerate}
\item\label{events}Quantum events axiom: the `states and observable
  axiom' (a) should be replaced by an axiom referring to events rather
  than systems, presumably by introducing a `Hilbert space for
  events'. This will hopefully permit a covariant formulation of the
  theory in which space and time are treated symmetrically.
\item\label{tp}Multiple events axiom: the tensor product structure of
  the `multiple systems axiom' (b) should be replaced by some
  mathematical construction that is able to describe multiple events.
  A quantum system should be introduced as a derivative notion, as a
  succession of quantum events.
\item\label{sch}Dynamics axiom: the privileged role of ``time'' in the
  `evolution axiom' (c) must be avoided. This might require a shift in
  the philosophy of the theory \cite{rov,wharton}, since quantum
  theory is formulated as a dynamical theory: given the initial
  conditions and given the dynamics (the Hamiltonian), the theory
  makes predictions or retrodictions {\em in time}.  General
  relativity in its covariant formulation\footnote{Of course, general
    relativity can also be given a non covariant formulation
    (geometrodynamics), and also a Hamiltonian formulation
    \cite{wald,mtw}.} is not a dynamical theory in the same sense
  \cite{rov}. At the
  very least, quantum mechanics should be modified to accommodate a
  multi-fingered time, e.g.~\cite{multif}, so that an operational
  observable meaning (proper time) can be attached to `time'.
\end{enumerate}
Comments on the above desiderata:\par\noindent (1)~A quantum theory
for events should be able to describe events that happen at a quantum
superposition of different times, and even quantum superpositions of
different causal orderings of events
\cite{causalset,fotini,bisio,switch,dribus,processmatrix}.  Conventional
quantum theory is unable to describe events in superposition
\cite{horw,horw1,greenberger}, although it seems possible to create
them experimentally \cite{palacios,lindner,paulus}. The possibility of
superposing events has the important implication that properties
incompatible to `spacetime position' exist. For example, the conjugate
property will be the event's energy-momentum, namely the event's
generator of spacetime translations.  This is a property that {\em
  must} be attached to a quantum event because of quantum
superposition and of complementarity. In the specific case where an
event refers to a transformation of a system's state, we can connect
the event's {\em time extent} to the system's energy through the
energy-time \cite{mandelstamm,margolus,qspeed} uncertainty relations.
In contrast, the position-momentum \cite{heisenberg,robertson}
uncertainty relation relates a system's spread in momentum to its {\em
  position uncertainty} (and not to its spatial extent). However, in
the case of more general notions of events, the physical
interpretation of the event's energy-momentum is unclear: in the
classical case, there is no energy or momentum necessarily associated
to an event, but one could assign to it the total energy-momentum of
the systems whose trajectories intersection defines the event.
Similarly, in the quantum case, in the spirit of relationalism of
``spacetime coincidences'' \cite{rov} an intriguing possibility is to
connect the energy-momentum of the event to the energy and momentum of
the (quantum) reference frame system employed to give a physical
significance to the quantum event.  In this respect, it has been shown
that defining time as ``what is shown on a clock'' and taking into
account the clock's energy, one can introduce a dynamical time
variable \cite{paw,ak,qtime}. A similar procedure can be implemented
for space \cite{kijowki,piron}, but it is not clear whether these
methods can be applied to spacetime \cite{futuro}.\par\noindent
(2)~The tensor product can be adapted to describe multiple events
\cite{pope,processmatrix,vedr}, but arguably it is not the natural way
to do it: the main reason for tensor products in quantum theory is
that they give the correct law of composition of probability
amplitudes for measurements on multiple systems, but they will not
give the correct composition law for measurements on multiple events,
because later events may be influenced by earlier ones. In particular,
a description of a quantum system as a ``succession of quantum
events'' would require properties that cannot be easily captured
through a tensor product because properties of a system at different
times cannot be accessed independently \cite{pope} (a measurement at
one time influences successive measurements). Even when there is no
causal connection, different observers can assign different temporal
ordering to spacelike events, so they will give a completely different
quantum account of their spacetime description \cite{peresmf}.
\par\noindent (3)~One may expect that the measurement problem is a
major drawback in the change of paradigm \cite{wharton} for quantum
mechanics in going from a dynamical theory that makes predictions
based on an initial state and on a dynamics to a theory that describes
the whole history of the system.  In truth, this problem can be
sidestepped by carefully avoiding any interpretation of the
``wavefunction collapse'' as a dynamical transformation
\cite{peresmf,rov} and by using von Neumann's description of a
measurement apparatus as a dynamical coupling of some memory degree of
freedom with the system being measured (\cite{von}, ch.~VI).  In this
way, a complete ``history state'' can be described, and the correct
measurement statistics and multiple-measurement correlations can be
recovered by simply using the Born rule as shown in \cite{qtime},
without having to invoke changes to the measurement axiom~(d).

A successful quantum theory for events would presumably be one that
satisfies the above requirements 1-3 and, concurrently, recovers the
conventional axiomatic formulation (a-d) in the appropriate limits,
e.g.~when defining a system as a succession of events. Promising and
clever theories that describe events in quantum mechanics have
appeared in the literature, for example
\cite{pope,automa,vedr,eeqt,bisio,processmatrix,stu}, but none of them
satisfy all these requirements. 

A motivation that it is quantum mechanics rather than general
relativity that may need reformulating comes from the universe's
expansion (more generically, from FLRW metrics). Indeed, the number of
events in today's universe is vastly larger than immediately after the
big bang. Nonetheless our current narrative in terms of systems rather
than events forces us to say that the system (i.e.~the number of
quantum degrees of freedom in the universe) is unchanged, since
Hilbert spaces cannot evolve\footnote{The isomorphism of Hilbert
  spaces of equal dimension implies that the Hilbert space of a
  quantum system is chosen {\em only} by matching its dimension to the
  system's number of degrees of freedom.}. So (i) a larger number of
events `happen' to the same number of systems. Similarly problematic
is black hole evaporation: since most of the Hawking radiation is
composed of low-energy photons, whether or not the initial information
is preserved, the process of creating a black hole from matter and
waiting for its evaporation ``creates'' degrees of freedom. With our
current theory, this can be described only by saying that (ii) these
degrees of freedom are only ``activated'' and were present from the
start in some previously existing system.  While the statements (i)
and (ii) may be enforced technically by appealing to the continuous
nature and to the infinite dimensionality of the Hilbert space of
quantum field theory, it is a rather silly narrative.
It would be much better to describe these phenomena in terms of a
variable number of events rather than in terms of fixed, eternal
systems. A further motivation for an event-based quantum mechanics is
the existence of vacuum solutions to Einstein's equations, namely
universes made only of spacetime, with no ``systems''.  Without a
prescription for quantum events, those universes cannot have a quantum
description (admittedly, this impossibility might be a feature: in
such universes no observers can presumably exist). If one were
successful in quantizing general relativity, arguably a good testbed
for it would be to provide a description of highly pathological
spacetimes, such as the G\"odel one \cite{goedel}.  Indeed, quantum
mechanics is unable to deal with closed timelike curves without a
profound modification \cite{deutsch,loop,looplong} and some proposed
approaches encounter difficulties if they are present, since in this
case the Hawking-Malament theorem \cite{malament} does not hold
\cite{dribus}.

In conclusion, general relativity and quantum mechanics are
incompatible also because of their current formulation. We have
suggested a possible roadmap 1-3 for a new quantum theory to overcome
this problem while still retaining the formulation (a-d) we all know
and love.

\vskip 1\baselineskip I acknowledge many fruitful discussions and a
longstanding collaboration with V.  Giovannetti and S. Lloyd. I
acknowledge funding from Unipv ``Blue sky'' project-grant
n.~BSR1718573 and the FQXi foundation grant FQXi-RFP-1513 ``the
physics of what happens''.


\begin{references}
\bibitem{unruhbook}W. Unruh, in Time's Arrows Today: Recent Physical
  and Philosophical Work on the Direction of Time, by Steven F.
  Savitt, Editor (Cambridge University Press, 1997).
\bibitem{qtime}V.Giovannetti, S.Lloyd, L.Maccone, Quantum time, Phys.
  Rev. D, {\bf 92}, 045033 (2015).
\bibitem{birrell}N.D. Birrell, P.C.W. Davies, Quantum fields in curved
  space (Cambridge Univ. Press, 1982).
\bibitem{mtw}C.W. Misner, K.S. Thorne, J.A. Wheeler, {Gravitation}
  (Freeman, 1973).
\bibitem{paw}D.N. Page, W.K. Wootters, Evolution without evolution:
  Dynamics described by stationary observables, Phys. Rev. D, 27, 2885
  (1983).
\bibitem{ak}Y. Aharonov, T. Kaufherr, Quantum frames of reference,
  Phys. Rev. D {\bf 30}, 368 (1984).
\bibitem{morse}P. McCord Morse, H. Feshbach, {Methods of Theoretical
    Physics, Part I} (McGraw-Hill, 1953), Chap. 2.6.
\bibitem{altri} T. Banks, TCP, quantum gravity, the cosmological
  constant and all that, Nucl. Phys. B {\bf 249}, 332 (1985); R.
  Brout, Found. Phys. {\bf 17}, 603 (1987); R. Brout, G. Horwitz, D.
  Weil, Phys. Lett. B {\bf 192}, 318 (1987); R. Brout, Z. Phys. B {\bf
    68}, 339 (1987); V. Vedral, Time, (Inverse) Temperature and
  Cosmological Inflation as Entanglement, arXiv:1408.6965 (2014); C.
  Marletto, V. Vedral, Evolution without evolution, and without
  ambiguities, Phys. Rev. D {\bf 95}, 043510 (2017); A.R.H. Smith, M.
  Ahmadi, Quantizing time: Interacting clocks and systems,
  arXiv:1712.00081 (2017).
\bibitem{rov}C. Rovelli, Quantum Gravity (Cambridge Monographs of
  Mathematical Physics, 2000).
\bibitem{ahaparad}Y. Aharonov, D. Rohrlich, Quantum Paradoxes: Quantum
  Theory for the Perplexed (Wiley-VCH, 2005).
\bibitem{wald}R. Wald, General Relativity, (Univ. Chicago Press,
  1984).
\bibitem{schutz}B. F. Schutz, A first course in general relativity
  (Cambridge Univ. Press, 1985).
\bibitem{wharton}K. Wharton, The Universe is not a Computer, in
  Questioning the Foundations of Physics, A. Aguirre, B. Foster and
  Z. Merali (Eds),; pp.177-190, Springer (2015), arXiv:1211.7081.
\bibitem{multif} P.A.M. Dirac, Relativistic Quantum Mechanics, Proc.
  R.  Soc. Lon. A {\bf 136}, 453 (1932). S. Tomonaga, On a
  Relativistically Invariant Formulation of the Quantum Theory of Wave
  Fields, Progr. Theor.  Phys. {\bf 1}, 27 (1946).
\bibitem{causalset} D.P. Rideout R.D. Sorkin, Classical sequential
  growth dynamics for causal sets Phys. Rev. D {\bf 61}, 024002
  (1999).  
\bibitem{fotini}F. Markopoulou, The internal description of a causal
  set, Commun. Math. Phys. {\bf 211}, 559 (2000).
\bibitem{switch} G. Chiribella, G. M. D'Ariano, P. Perinotti, B.
  Valiron, Quantum computations without definite causal structure,
  Phys. Rev. A {\bf 88}, 022318 (2013).
\bibitem{bisio}A. Bisio, G. Chiribella, G.M. D'Ariano, P. Perinotti,
  Quantum Networks: general theory and applications, Acta Physica
  Slovaca {\bf 61}, 273 (2011), arXiv:1601.04864.
\bibitem{dribus} B.F. Dribus, Discrete Causal Theory (Springer, 2017),
  ch.~2.7-2.8.
\bibitem{processmatrix} O. Oreshkov, F. Costa. C.  Brukner, Quantum
  correlations with no causal order, Nature Comm. {\bf 3}, 1092
  (2012); M. Ara\'ujo, A. Feix, M.  Navascu\'es, C.  Brukner, A
  purification postulate for quantum mechanics with indefinite causal
  order, Quantum {\bf 1}, 10 (2017).
\bibitem{horw} L.P. Horwitz, {Quantum interference in time},
  Found. Phys. {\bf 37,} 734 (2004).
\bibitem{horw1} L.P. Horwitz, {On the significance  of a recent
    experimentally demonstrating quantum interference in time},
  Phys. Lett. A {\bf 355}, 1 (2006).
\bibitem{greenberger}D.M. Greenberger, Conceptual Problems Related to
  Time and Mass in Quantum Theory, arXiv:1011.3709 (2010).
\bibitem{palacios}A. Palacios, T.N. Rescigno, C.W. McCurdy,
  Two-Electron Time-Delay Interference in Atomic Double Ionization
  by Attosecond Pulses, Phys. Rev. Lett. {\bf 103}, 253001 (2009).
\bibitem{lindner} F. Lindner, M.G. Sch\"atzel, H. Walther, A.
  Baltuska, E. Goulielmakis, F. Krausz, D.B. Milosevi\'c, D. Bauer, W.
  Becker, and G.G. Paulus, Attosecond Double-Slit Experiment,
  Phys. Rev. Lett. {\bf 95}, 040401 (2005).
\bibitem{paulus} G.G. Paulus, F. Lindner, H. Walther, A. Baltuska, E.
  Goulielmakis, M. Lezius, and F. Krausz, Measurement of the Phase of
  Few-Cycle Laser Pulses, Phys. Rev. Lett. {\bf 91}, 253004 (2003).
\bibitem{mandelstamm} L. Mandelstam and I. G. Tamm, The uncertainty
  relation between energy and time in nonrelativistic quantum
  mechanics, J. Phys. USSR {\bf 9}, 249 (1945).
\bibitem{margolus}N. Margolus and L. B. Levitin, The maximum speed of
  dynamical evolution, {Physica D} {\bf 120}, 188 (1998).
\bibitem{qspeed} V. Giovannetti, S. Lloyd, L. Maccone, Quantum limits
  to dynamical evolution, Phys. Rev. A {\bf 67}, 052109 (2003).
\bibitem{heisenberg} W. Heisenberg, \"Uber den anschaulichen Inhalt
  der quantentheoretischen Kinematik und Mechanik, Zeit. Phys. {\bf
    43}, 172 (1927), English translation in \cite{zur}, pg.  62--84.
\bibitem{robertson}H. P. Robertson, {The uncertainty principle}, Phys.
  Rev. {\bf 34}, 163 (1929).
\bibitem{kijowki} J. Kijowski, {On the time operator in quantum
    mechanics and the heisenberg uncertainty relation for energy and
    time} Rep. Math. Phys. {\bf 6}, 361 (1974).
\bibitem{piron}C. Piron, { Un nouveau principe d'\'evolution
    r\'eversible et une g\'eneralisation de l'equation de
    Schroedinger,} C.R. Acad. Seances (Paris) A {\bf 286}, 713 (1978).
\bibitem{futuro}V. Giovannetti, S. Lloyd, L. Maccone, Quantum
  spacetime from constrains, in preparation.
\bibitem{pope}Y. Aharonov, S. Popescu, J. Tollaksen, in {Quantum
    Theory: A Two-Time Success Story}, Chap.~3, pp 21-36 (2014),
  arXiv:1305.1615 [quant-ph].
\bibitem{vedr}J.F. Fitzsimons, J.A. Jones, V. Vedral, Quantum
  correlations which imply causation. Sci. Rep. {\bf 5}, 18281 (2015).
\bibitem{peresmf}A.~Peres, Classical interventions in quantum systems.
  II. Relativistic invariance Phys. Rev. A {\bf 61}, 022117 (2000).
\bibitem{von} J.  von Neumann, {Mathematical Foundations of
    Quantum Mechanics} (Princeton Univ. Press, 1955).
\bibitem{automa} G.M. D'Ariano, P. Perinotti, Quantum cellular
  automata and free quantum field theory, Front. Phys. {\bf 12}(1),
  120301 (2017), arXiv:1608.02004.
\bibitem{eeqt}P. Blanchard, A. Jadczyk, Event-enhanced quantum theory
  and piecewise deterministic dynamics, Ann. Physik {\bf 4}, 583
  (1995).
\bibitem{stu} E.C.G. Stueckelberg, La signification du temps propre
  en m\'ecanique ondulatoire, Helv. Phys. Acta {\bf 14,} 322 (1941);
  E.C.G. Stueckelberg, Remarque \`a propos de la cr\'eation de
  paires de particules en th\'eorie de relativit\'e, Helv. Phys.
  Acta {\bf 14,} 588 (1941); E.C.G. Stueckelberg, La m\'ecanique du
  point mat\'eriel en th\'eorie des quanta, Helv. Phys. Acta {\bf
    15,} 23 (1942).
\bibitem{goedel} K. G\"{o}del, An Example of a New Type of
  Cosmological Solutions of Einstein's Field Equations of Gravitation,
  Rev.  Mod. Phys. {\bf 21}, 447 (1949).
\bibitem{deutsch}D. Deutsch, Quantum mechanics near closed timelike
  lines, Phys. Rev. D {\bf 44}, 3197 (1991).
\bibitem{loop} S.~Lloyd, L.~Maccone, R.~Garcia-Patron, V.~Giovannetti,
  Y.~Shikano, S.~Pirandola, L.A.~Rozema, A.~Darabi, Y.~Soudagar,
  L.K.~Shalm, A.M.~Steinberg, Closed timelike curves via
  postselection: theory and experimental test of consistency, Phys.
  Rev.  Lett. {\bf 106}, 040403 (2011).
\bibitem{looplong} S. Lloyd, L. Maccone, R. Garcia-Patron, V.
  Giovannetti, Y. Shikano, The quantum mechanics of time travel
  through post-selected teleportation, Phys. Rev. D {\bf 84}, 025007
  (2011).
\bibitem{malament}D.B. Malament, The class of continuous timelike
  curves determines the topology of spacetime, J. Math. Phys. {\bf
    18}, 1399 (1977).
\bibitem{zur} J. A. Wheeler, H. Zurek, {Quantum Theory and
    Measurement}, (Princeton Univ.~Press, Princeton, 1983).
\bibitem{diracsch}P.A.M. Dirac, Foundations of Quantum Mechanics,
  Nature, {\bf 203}, 115 (1964).
\bibitem{franson}J.D. Franson, Velocity-dependent forces, Maxwell's
  demon, and the quantum theory, arXiv:1707.08059 (2017).
\end{references}
\end{document}